\documentclass[useAMS,usenatbib]{mn2e}
\usepackage{graphicx}
\usepackage{times}

\def \aap{A\&A}

\def \aj{AJ}

\def \apjl{ApJ}
\def \apjs{ApJS}
\def \apj{ApJ}

\def \asp{Astron.~Soc.~Pac.}

\def \mnras{MNRAS}

\def \prl{Phys.~Rev.~Lett.}

\newcommand{\CIV}{\hbox{{\rm C}{\sc \,iv}}}

\newcommand{\FeII}{\hbox{{\rm Fe}{\sc \,ii}}}

\newcommand{\MgI}{\hbox{{\rm Mg}{\sc \,i}}}
\newcommand{\MgII}{\hbox{{\rm Mg}{\sc \,ii}}}

\newcommand{\HI}{\hbox{{\rm H}{\sc \,i}}}

\newcommand{\lya}{\hbox{{\rm Ly}$\alpha$}}

\newcommand{\msun}{M$_{\odot}$}

\newcommand{\NHI}{\hbox{$N_{\HI}$}}
\newcommand{\ma}{\hbox{$\lambda 2796$}}
\newcommand{\mb}{\hbox{$\lambda 2803$}}
\newcommand{\mc}{\hbox{$\lambda 2852$}}
\newcommand{\fa}{\hbox{$\lambda 2600$}}
\newcommand{\LL}{\hbox{$\lambda\lambda$}}
\newcommand{\zab}{\hbox{$z_{\rm abs}$}}
\newcommand{\kms}{\hbox{${\rm km\,s}^{-1}$}}
\newcommand{\nn}{\nonumber}
\newcommand{\dNdz}{\hbox{$\frac{\mathrm d N}{\mathrm d z}$}}
\newcommand{\dNdl}{\hbox{$\frac{\mathrm d N}{\mathrm d l}$}}
\newcommand{\Hgamma}{\hbox{$\Gamma(\frac{1}{2})\Gamma(\frac{\gamma-1}{2})/\Gamma(\frac{\gamma}{2})$}}

\newcommand{\ngal}{\hbox{33,348}}
\newcommand{\ngalsim}{\hbox{20,000}}
\newcommand{\ngalused}{\hbox{19,496}}
\newcommand{\Aauto}{\hbox{$0.175\pm 0.028$}}
\newcommand{\Bauto}{\hbox{$-0.780\pm 0.147$}}
\newcommand{\Across}{\hbox{$0.179\pm 0.026$}}
\newcommand{\Bcross}{\hbox{$-0.955\pm 0.118$}}
\newcommand{\Arel}{\hbox{$0.84 \pm 0.09$}}
\newcommand{\Arelcorr}{\hbox{$0.67 \pm 0.09$}}
\newcommand{\Arelcorrsys}{\hbox{$0.67 \pm 0.07\pm {0.05}$}}
\newcommand{\rnot}{\hbox{$6.2^{+1.1}_{-1.0}$}}
\newcommand{\LRGminmass}{\hbox{$3.5\times10^{12}$}}
\newcommand{\LRGmassrangemin}{$\sim2$--$8\times 10^{11}$\,\msun}
\newcommand{\LRGmassrange}{$0.5$--$2.5\times 10^{12}$\,\msun}
\newcommand{\overestimate}{\hbox{$25\pm10$}}

\def\bsp_small{\vspace{0.5cm}\small\noindent This paper
has been typeset from a \TeX / \LaTeX\ file prepared by the author.}

\title[The clustering of LRGs around \MgII\ absorbers]{The clustering of
  Luminous Red Galaxies around $\bmath{\MgII}$ absorbers}
  \author[N. Bouch\'e, M. T. Murphy, C. P\'eroux]{Nicolas
  Bouch\'e$^1$\thanks{E-mail: nbouche@eso.org (NB); mim@ast.cam.ac.uk
  (MTM); cperoux@eso.org (CP)}, Michael T.~Murphy$^2$\footnotemark[1],
  C\'eline P\'eroux$^1$\footnotemark[1]\\ $^1$European Southern
  Observatory, Karl-Schwarzschild-Str 2, D-85748 Garching, Germany\\
  $^2$Institute of Astronomy, University of Cambridge, Madingley Road,
  Cambridge CB3 0HA, UK}

\begin{document}

\date{Accepted 2004 August 25.
      Received 2004 May 31 ;
      in original form 2004 September 20}

\pagerange{\pageref{firstpage}--\pageref{lastpage}}
\pubyear{2004}

\maketitle

\label{firstpage}

\begin{abstract}
We study the cross-correlation between 212 \MgII\ quasar absorption systems
and $\sim\!\ngalsim$ Luminous Red Galaxies (LRGs) selected from the Sloan
Digital Sky Survey Data Release 1 in the redshift range
$0.4\!\leq\!z\!\leq\!0.8$. The \MgII\ systems were selected to have \LL
2796 \& 2803 rest-frame equivalent widths $\geq\!1.0$\,\AA\ and
identifications confirmed by the \FeII\ \fa\ or \MgI\ \mc\ lines. Over
comoving scales 0.05--13$h^{-1}{\rm \,Mpc}$, the \MgII--LRG
cross-correlation has an amplitude \Arelcorr\ times that of the LRG--LRG
auto-correlation. Since LRGs have halo-masses greater than
\LRGminmass\,\msun\ for $M_R\!\la\!-21$, this relative amplitude
implies that the absorber host-galaxies have halo-masses greater than
\LRGmassrangemin. For $10^{13}$\,\msun\ LRGs, the absorber host-galaxies have halo-masses
\LRGmassrange. Our results appear consistent with those of \citet{SteidelC_94a} 
who found that \MgII\ absorbers with $W_{\rm
r}^{\rm MgII}\!\geq\!0.3$\,\AA\ are associated with $\sim\!0.7L_B^*$  
galaxies.
\end{abstract}

\begin{keywords}
cosmology: observations --- galaxies: evolution ---  galaxies: halos ---
 quasars: absorption lines
\end{keywords}

\section{Introduction}

The connection between quasar (QSO) absorption line systems and galaxies
\citep{BergeronJ_91a} is important to our understanding of galaxy
evolution. Absorption lines provide detailed information about the physical
conditions and kinematics of galaxies out to large impact parameters
($R\!\ga\!100{\rm \,kpc}$), regardless of the absorber's intrinsic
luminosity \citep[e.g.][]{RauchM_96a,EllisonS_00a}. \MgII\ \LL 2796 \& 2803
are amoungst the most studied metal lines since the doublet signature makes
for easy detection.

Past results show that \MgII\ absorbers are not unbiased tracers of
galaxies but are biased towards late-type galaxies which do not evolve
strongly from $z\!\simeq\!1$. The morphological constraints come from
imaging by \citet{SteidelC_92a} and \citet*{SteidelC_94a} who found that
\MgII\ absorber host-galaxies have $K$-band luminosities consistent with
normal $0.7L^*_B$ Sb galaxies. Further Hubble Space Telescope imaging
\citep{SteidelC_98a} indicated that \MgII\ absorbers at $z_{\rm abs}\!<\!1$
are drawn from field galaxies of all disk morphological types. These
galaxies are also found to have roughly constant star formation rate since
$z\!\sim\!1$: from a sample of 58 \MgII\ absorbers with rest-frame
equivalent widths $W_{\rm r}^{\rm MgII}\!\geq\!0.30$\,\AA, \citet{SteidelC_94a}
found that the mean rest-frame $M_B$ and $B\!-\!K$ colours of the
host-galaxies do not evolve in the redshift range $0.2\!<\!z_{\rm
abs}\!<\!1$. Furthermore, their rest-frame $K$-band luminosity function
(LF) closely matches the $K$-band LF at $z\!=\!0$ down to $0.05L^*_K$. This
`no evolution' picture of the $K$-band LF implies little or no stellar mass
evolution from $z\!\simeq\!1$. However, using more direct constraints from
deep imaging surveys, \citet{DickinsonM_03a} and \citet{RudnickG_03a} find
that the stellar mass increased by a factor of 2 over this epoch. Thus,
these earlier studies imply that \MgII\ absorption-selected galaxies are
biased towards non-evolving, luminous disk morphological types.

From the observed absorber--host-galaxy impact parameter distribution,
\citet{SteidelC_93a,SteidelC_95b} constrained the cross-section of \MgII\
absorbers with $W_{\rm r}^{\rm MgII}\!\geq\!0.30$\,\AA\ to have radius
$R_\times\!\sim\!40h^{-1}{\rm \,kpc}$ (physical, $70h^{-1}{\rm \,kpc}$
comoving). In addition, these systems are always found to be associated
with neutral hydrogen absorbers in the Lyman limit regime
($\log\NHI\!\geq\!3\!\times\!10^{17}{\rm \,cm}^{-2}$).

Little information currently exists about the environment of \MgII\
absorbers on scales up to $10h^{-1}{\rm \,Mpc}$. Recently,
\citet*{HainesC_04a} analysed the clustering of early-type galaxies around
two \MgII\ absorbers at $z_{\rm abs}\!=\!0.8$ \& $1.2$ using wide field
images ($40\arcmin \times 35\arcmin$). They find a significant excess of
galaxies across the field and conclude that large-scale structures
containing \MgII\ absorbers mark out volumes of enhanced galaxy density.

All the above studies are based on relatively small ($\la\!50$) samples of
\MgII\ absorbers and, with the exception of \citet{HainesC_04a}, relatively
small-area ($\la\!10\arcmin \times 10\arcmin$) galaxy surveys. The Sloan
Digital Sky Survey \citep[SDSS;][]{StoughtonC_02a} allows us to
significantly transcend these limitations. In this paper we use
$\sim\!\ngalsim$ Luminous Red Galaxies (LRGs) from SDSS Data Release 1
\citep[DR1;][]{AbazajianK_03a,StraussM_02a} to constrain the environment,
and more specifically, the mass of the halos associated with 212 \MgII\
absorbers. In a hierarchical galaxy formation scenario, the amplitude ratio
of the \MgII--LRG cross-correlation and LRG--LRG auto-correlation is a
measure of the relative masses of the halos associated with \MgII\
absorbers and LRGs.

Throughout this paper, we adopt $\Omega_{\rm M}\!=\!0.3$,
$\Omega_\Lambda\!=\!0.7$, $\sigma_8=0.8$, and $H_0\!=\!100 h\,\kms{\rm \,Mpc}^{-1}$. Thus,
at $z\!=\!0.6$, $1\arcsec$ corresponds to $7.44h^{-1}{\rm \,kpc}$ and
$1\arcmin$ corresponds to $446h^{-1}{\rm \,kpc}$, both {\it comoving}. At
that redshift $H(z)\!=\!1.39H_0$, so $\delta z\!=\!0.1$ corresponds to
$216 h^{-1}{\rm \,Mpc}$ in comoving coordinates.
 
\section{Sample definitions}

\begin{figure}
\centerline{\includegraphics[height=83mm,angle=270]{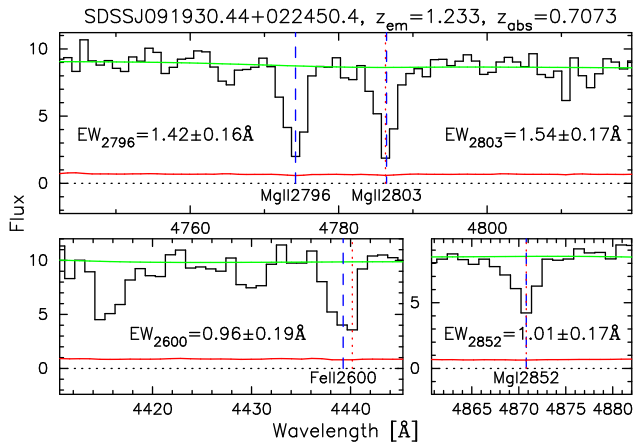}}
\caption{Example \MgII\ absorption system showing the \MgII\ \LL 2796/2803
  doublet and the supporting transitions, \FeII\ \fa\ and \MgI\ \mc. Note
  also the \FeII~$\lambda$2587 line at $\lambda_{\rm obs}\approx
  4415$\AA. The upper solid line is our fitted continuum and the lower
  solid line is the $1\,\sigma$ error array.}
\label{fig:example}
\end{figure}

\subsection{$\bmath{\MgII}$ absorbers}

For the 16713 QSO spectra in the SDSS QSO sample of \citet{SchneiderD_03a},
we searched for \MgII\ \LL 2796/2803 absorption doublets with $\zab \le
0.8$ using a largely automated technique. A third-order polynomial was
fitted to overlapping 2500-\kms\ sections of each QSO spectrum from
10,000\,\kms\ above the \lya\ emission line to 10,000\,\kms\ below the
\MgII\ emission line. Pixels with flux $>\!2\,\sigma$ below and
$>\!5\,\sigma$ above the continuum are rejected and the continuum is
re-fitted to the remaining points. This process is iterated until no more
points are rejected. Overlapping portions of adjacent continua are joined
by weighting each linearly from zero at the edge to unity at the
centre. The final continuum is smoothed over 11 pixels
($\approx\!760$\,\kms). The 2500\,\kms\ chunk-size is small enough to fit
most emission features but large enough so that strong \MgII\ doublets do
not cause significant spurious bends in the continuum.

Candidate \MgII\ \ma\ lines are searched for by identifying the pixel within a
$\Delta\lambda_{\rm rest}\!=\!7$\,\AA\ sliding window with the most
significant flux deviation $>\!2\,\sigma$ below the continuum. This
$\Delta\lambda_{\rm rest}$ accounts for most of the \MgII~\ma\ absorption
while avoiding significant overlap with the \MgII~\mb\ line. A similar
window is centred on the \MgII~\mb\ wavelength with the same redshift,
\zab, as the putative \MgII~\ma\ line. If $W_{\rm r}^{\rm
MgII}\!\geq\!1.0$\,\AA\ and the mean signal-to-noise ratio (S/N) is
$\geq\!10$ within each window, the system is identified as a candidate
\MgII\ absorber.

Candidates require at least one supporting transition to be considered real
detections: \FeII~\fa\ or \MgI~\mc. \FeII\ is the preferred line since it
is stronger than \MgI. \FeII~\fa\ is, in principle, detectable in SDSS
spectra when $\zab\!\ga\!0.47$ (i.e.~$\lambda_{\rm obs}\!\ga\!3800$\AA) and
so we use the criterion advocated by \citet{NestorD_03a} to select for
damped \lya\ systems (DLAs), $W_{\rm r}^{\rm FeII}\!\geq\!0.5$\AA. If
\FeII\ \fa\ is not detectable or if ${\rm S/N}\!<\!10$ in the \FeII\ \fa\
window, we require that \MgI\ \mc\ is detected with ${\rm S/N}\!\geq\!10$
and ${W_{\rm r}^{\rm MgI}\!\geq\!0.2}$\AA. Several caveats apply to the
above requirements, particularly when one or more transitions fall near
emission features and the broad absorption features often associated with
them. We will describe these caveats in detail in a later paper, suffice it
to note here that they apply to $\la\!20$\,per cent of (real) systems.

Finally, we remove clearly spurious candidates by visually inspecting each
\MgII\ spectrum. The most common mis-identification is broad \CIV\
absorption near the \CIV\ emission line.

With the above algorithm we detected and visually confirmed 212 \MgII\
absorbers in the DR1. A typical \MgII\ absorption doublet is shown in
Fig.~\ref{fig:example}. Note that, in addition to the crucial lines for
selection, the \FeII\ $\lambda$2587 line also often confirms the
detection. We provide a catalogue of the \MgII\ absorbers in Table
\ref{table:mg} and the absorption redshift distribution is shown in
Fig.~\ref{fig:zspec}.

\begin{table}
\begin{center}
\vspace{-1mm}
\caption{Catalogue of 212 \MgII\ absorbers from the SDSS DR1 with
$0.4\!\leq\!z_{\rm abs}\!\leq\!0.8$. The J2000 name, QSO and absorption
redshifts and the measured $W_{\rm r}$ for \MgII \LL 2796 \& 2803, \MgI\
\mc\ and \FeII\ \fa\ are given. Here we show only a small sample from the
full table which is available in the electronic edition of this paper and
from http://www.ast.cam.ac.uk/$\sim$mim/pub.html. Full name designations
and statistical uncertainties in $W_{\rm r}$ are given in the electronic
version.}
\vspace{-4mm}
\label{table:mg}
\begin{tabular}{lcccccc}\hline
 & & &\multicolumn{4}{c}{Rest equivalent width $\left[{\rm \AA}\right]$}\\
\multicolumn{1}{c}{SDSSJ}&\multicolumn{1}{c}{$z_{\rm qso}$}&\multicolumn{1}{c}{$z_{\rm abs}$}&\multicolumn{1}{c}{2796}&\multicolumn{1}{c}{2803}&\multicolumn{1}{c}{2852}&\multicolumn{1}{c}{2600}\\\hline
004041$-$005537 & 2.090 & 0.6612 & 4.37 & 3.46 &  0.86 & 1.83\\
004721$+$154652 & 1.272 & 0.7742 & 1.31 & 1.92 &  0.57 & 1.12\\
005130$+$004150 & 1.190 & 0.7394 & 2.19 & 1.71 &  0.58 & 1.27\\
005408$-$094638 & 2.128 & 0.4778 & 2.18 & 2.09 &  1.20 & 2.17\\\hline
\end{tabular}
\end{center}
\end{table}

\subsection{Luminous red galaxies}\label{section:LRGs}

SDSS DR1 contains more than $10^6$ LRGs over $\sim\!2000$ sq.~degrees which
have luminosities $M_g\!<\!-21$ and fall on the red sequence with
$(u\!-\!g)_0\!\simeq\!2$ \citep{EisensteinD_01a,ScrantonR_04a}.

\begin{figure}
\centerline{\includegraphics[width=70mm]{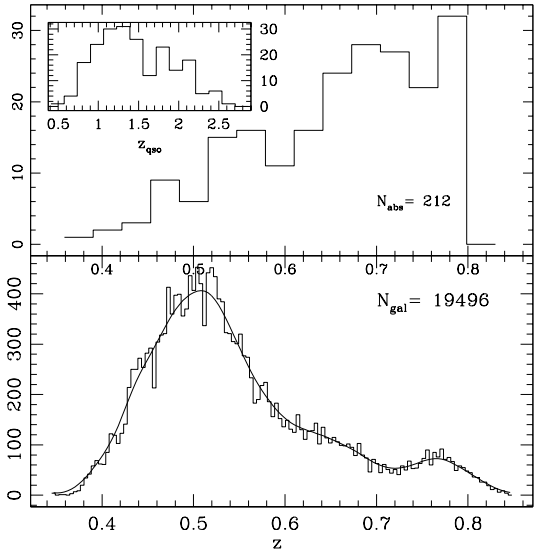}}
\caption{Redshift distribution of the LRGs (bottom) and \MgII\ systems
(top). The solid curve shows the LRG distribution smoothed with a Gaussian
kernel. The inset shows the distribution of QSO emission redshifts.}
\label{fig:zspec}
\end{figure}

For each \MgII\ absorber, galaxies meeting the following criteria were
extracted from the SDSS DR1 galaxy catalogue:
\begin{eqnarray}
i^*_{\rm petro}&<&21,\label{LRG:selection1}\\
0.7(g^*-r^*)+1.2[(r^*-i^*)-0.18] &>& 1.6\,, \label{LRG:scranton1}\\
(g^*-r^*) &>& 1\,, \label{LRG:scranton2} \\
({\rm d}_\perp \equiv) (r^*-i^*)-(g^*-r^*)/8 &>& 0.4\,, \label{LRG:scranton3} \\
r^*_{\rm psf}-r^*_{\rm model}&>&0.3\,,  \label{LRG:selection2}\\ 
\left|z_{\rm phot}-z_{\rm abs}\right|&<&0.05\,. \label{LRG:selection3}
\end{eqnarray}
We also required errors on the model magnitudes to be less than $0.2{\rm
\,mag}$ in $r^*$ and $i^*$, and we excluded objects flagged by SDSS as
BRIGHT, SATURATED, MAYBE\_CR or EDGE.  The model magnitudes were used to
compute the colours. Equations
(\ref{LRG:selection1})--(\ref{LRG:scranton3}) are the LRG selection
criteria of \citet{ScrantonR_04a}. Criterion \ref{LRG:scranton3} is
equivalent to imposing $z_{\rm phot}\!\ga\!0.3$.  Criterion
\ref{LRG:selection2} separates stars from galaxies.  Criterion
\ref{LRG:selection3} is the selection of galaxies within a redshift slice
of width $W_z\!=\!0.1$ around $z_{\rm abs}$ using the photometric
redshifts, $z_{\rm phot}$, of \citet{CsabaiI_03a} who showed these to be
accurate to $\sigma_z\!=\!0.1$ at $r'\!<\!21$. The choice of the slice
width $W_z\!=\!0.1$ corresponds to $\sim\!200h^{-1}{\rm \,Mpc}$ (co-moving)
and is arbitrary. It is a compromise to optimise the signal-to-noise: too
small a width will yield too few correlated pairs, too large a width will
wash out the signal. Finally, we remove the 10\,per cent of the galaxies
with problematic photometric redshifts by requiring that galaxies have
$z_{\rm phot}$ uncertainties $\sigma_{z_{\rm phot}}\!<\!0.5$. A total of
\ngal\ galaxies met all these criteria in our 212 fields
($\sim\!300$\,sq.~degrees). Fig.~\ref{fig:zspec} shows the redshift
distribution of these LRGs for the 212 fields.  We used the spectroscopic
redshift when available, which includes only $\sim\!160$ LRGs. This
situation will change in the future with the 2dF/SDSS program to
obtain spectra of LRGs. 

LRGs are expected to have halo-masses
$>\!10^{12}$\,\msun. \citet{BrownM_03a} showed that the clustering of red
($B_W\!-\!R\!>\!1.44$) galaxies between $z\!=\!0.3$ and $0.9$ in the NOAO
deep wide survey is a strong function of luminosity: in the luminosity
range $-21.5\!<\!M_R\!<\!-20.5$, the correlation length is $6.3 \pm 0.5
h^{-1}{\rm \,Mpc}$, and rapidly increases to $11.2h^{-1}{\rm \,Mpc}$ at
$M_R\!=\!-22$.  Such strong clustering is consistent with halo-masses of
$3.5\times 10^{12}$ to $3\times 10^{13}$\,\msun\ using the bias
prescription of \citet{MoH_02a}.

\vspace{-0.5cm}

\section{Results}
\label{section:results}

\subsection{Theoretical background}
\label{section:theory}

A widely used statistic to measure the clustering of galaxies is the
correlation function, $\xi(r)$. The absorber--galaxy cross-correlation,
$\xi_{\rm ag}$, is defined from the conditional probability of finding a
galaxy in a volume d$V$ at a distance $ r=|\mathbf r_2\!-\!\mathbf r_1|$,
given that there is a \MgII\ absorber at $\mathbf r_1$:
\begin{equation}
P({\rm LRG}|\MgII) =\overline n_u \left[1+\xi_{\rm ag}(r) \right] \mathrm d
V,
\label{eq:cross}
\end{equation}
where $\overline n_u$ is the unconditional background galaxy density.

The observed amplitudes of the auto- and cross-correlation functions
are related to the dark matter correlation function, $\xi_{\rm DM}$,
through the bias, $b(M)$, which is a function of the dark matter halo-mass
\citep[e.g.][]{MoH_93a,MoH_02a}:
\begin{eqnarray}
\xi_{\rm gg}(r)&=&b^2(M_{\rm g})\,\xi_{\rm DM}(r)\,, \label{eq:biasCDM}\\
\xi_{\rm ag}(r)&=&b(M_{\rm a})\,b(M_{\rm g})\,\xi_{\rm DM}(r)\,. \label{eq:biascross}
\end{eqnarray}
Thus, the  amplitude ratio of the cross- to auto-correlation,
which is $(r_{\rm 0,ag}/r_{\rm 0,gg})^{\gamma}$ for $\xi(r)=(r/r_{0})^{-\gamma}$,
  is a measurement of the bias ratio $b(M_{\rm a})/b(M_{\rm g})$
 which in turn yields  the relative halo-masses ($M_{\rm a}/M_{\rm g}$). 
This assumes that  $\xi_{\rm ag}$ and $\xi_{\rm gg}$ have the same slope $\gamma$.

Since our LRG sample is made up of galaxies with photometric redshifts, we
computed the projected cross- and auto-correlation functions, i.e.
as a function of physical distance $r_{\theta}=D_A(1+z)\theta$ in comoving Mpc
with  $D_A$   the angular diameter distance.
From the definitions of $w_{\rm gg}(r_\theta)$ \citep[e.g.][]{PhillippsS_78a,PeeblesP_93a}
 and $w_{\rm ag}(r_\theta)$ \citep[e.g.][]{EisensteinD_03a,AdelbergerK_03a}, 
 the amplitude of both $w_{\rm gg}(r_\theta)$ and $w_{\rm ag}(r_\theta)$
is inversely proportional to   $1/W_z$, where $W_z$ is the width of
redshift distribution $\dNdz$. For a top-hat  $\dNdz$,
the ratio
$w_{\rm ag}(r_\theta)/w_{\rm ag}(r_\theta)$ is exactly the bias ratio $b(M_{\rm a})/b(M_{\rm g})$,
irrespective of  $W_z$~\footnote{This result is demonstrated in Appendix A.}.
 In the case of a Gaussian redshift distribution $\dNdz$, 
$w_{\rm ag}(r_\theta)/w_{\rm ag}(r_\theta)$ is overestimated by \overestimate\ per cent.
This factor was determined using (i) numerical integration and (ii)
mock catalogues \citep[from the GIF2 collaboration,][]{GaoL_04a}
made of galaxies that had a 
redshift uncertainty equal to the slice width, $W_z$, as in the case 
of our   LRG sample. Note that this factor depends   on the shape of $\dNdl$, 
not its width.

\subsection{$\bmath{\MgII}$--luminous red galaxy cross-correlation}
\label{section:crosscorr}

\begin{figure}
\centerline{\includegraphics[width=80mm]{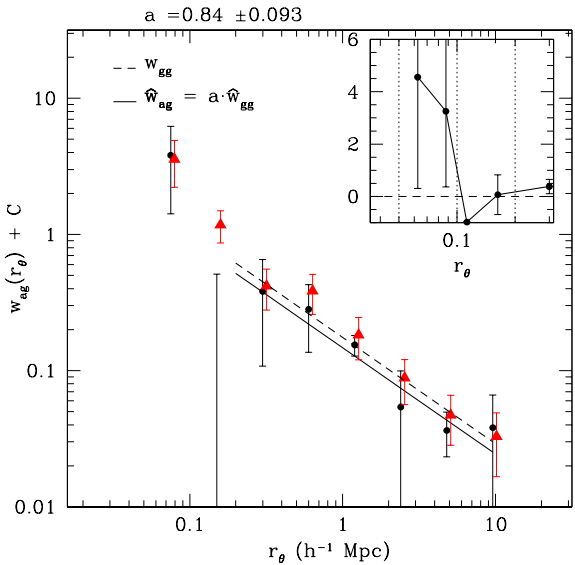}}
\caption{The projected \MgII--LRG cross-correlation, $w_{\rm
ag}(r_\theta)$, for 212 \MgII\ absorbers and \ngalused\ LRGs is shown by
the filled circles.  The error bars are computed using the Jackknife
technique. The filled triangles show the LRG--LRG auto-correlation, $w_{\rm gg}$, 
offset by 0.025 in $r_\theta$ for clarity, 
and the dashed line shows the fit to $w_{\rm gg}$. The solid line shows the
relative amplitude of $w_{\rm ag}$ and $w_{\rm gg}$, i.e.~using $\hat
w_{\rm ag}=a\times \hat w_{\rm gg}$ for scales $r_\theta\!>\!200h^{-1}{\rm
\,kpc}$ since the smallest scales will be affected by the finite
cross-section of the absorbers. The best-fitting relative amplitude is
$a\!=\!\Arel$. The inset shows the first two bins, 50--100$h^{-1}{\rm
\,kpc}$ and $100$--$200h^{-1}{\rm \,kpc}$ (dotted lines), split into two
sub-bins. The third sub-bin, 100--150$h^{-1}{\rm \,kpc}$, contains no
galaxies due to the absorber cross-section effects.}
\label{fig:xcorr:auto}
\end{figure}

Fig.~\ref{fig:xcorr:auto} (filled circles) shows $w_{\rm ag}$ for the entire
sample, where we used the following estimator of $w_{\rm ag}(r_\theta)$
\citep[also advocated by][]{AdelbergerK_03a}:
\begin{equation}
1+ w_{\rm ag}(r_\theta) = \frac{\rm AG}{\rm AR}\,, \label{eq:estimator}
\end{equation}
where AG is the observed number of absorber--galaxy pairs between
$r_\theta-dr/2$ and $r_\theta+\mathrm d r/2$, summed over all the fields.
AR is the normalized absorber--random galaxy pairs where the normalization is
applied to each field independently: ${\rm AR}\!=\!\sum_i {\rm AR}^i \;
N_g^i/N_r^i$, where AR$^i$ is the number of random pairs in field $i$, and
$N_g^i$ ($N_r^i$) is the total number of galaxies (random galaxies) for
that field.  From the sample of \ngal\ objects within the initial search
radius of $40\arcmin$, there are \ngalused\ objects within
$r_\theta\!=\!12.8h^{-1}{\rm \,Mpc}$ which is the outer radius of the
largest bin used. Taking into account the areas missing from the SDSS
within our search radius, we generated approximately $200$ times more random
galaxies to reduce the shot noise of AR to an insignificant proportion.
 Table \ref{table:pairs} shows the total number of
pairs, AG, and the expected number of pairs, AR, if \MgII\ absorbers and
LRGs were not correlated.

The $w_{\rm ag}$ error bars are computed using the jackknife estimator
\citep{EfronB_82a}: we divide the sample into 10 parts and compute the
covariance matrix from the $N_{\rm jack}\!=\!10$ realisations for each
part:
\begin{equation}
{\rm COV}_{ij}=\frac{N_{\rm jack}-1}{N_{\rm jack}}\sum_{k=1}^{N_{\rm jack}}  [w_k(r_{\theta_i})-\overline w(r_{\theta_i})]\cdot [w_k(r_{\theta_j})-\overline w(r_{\theta_j})]
\label{eq:covariance}
\end{equation}
where $w_k$ is the $k$th measurement of the cross-correlation and
$\overline w$ is the average of the $N_{\rm jack}$ measurements.

\begin{table}
\begin{center}
\vspace{-1mm}
\caption{The total number of absorber--LRG pairs AG and the number of
absorber--random pairs AR, for the cross-correlation shown in
Fig.~\ref{fig:xcorr:auto}. The number of pairs expected if one extrapolated
$w_{\rm ag}$ to $r_\theta\!<\!0.2h^{-1}{\rm kpc}$ are shown in parentheses (see
text).}
\label{table:pairs}
\vspace{-1mm}
\begin{tabular}{ccccc}\hline
	{$r_\theta$} & {AG} & {AR} & {$w_{\rm ag}$} & {$\sigma(w)$} \\
	{[$h^{-1}{\rm \,Mpc}$]} & { } & { } & { } & { }   \\\hline
 0.05--0.1  &   4(2.5)      &  0.834   &     3.797   & 2.40   \\
 0.1--0.2   &   3(7)       &  3.45    &  $-$0.135   & 0.625  \\
 0.2--0.4   &   20    &  14.7    &     0.361   & 0.274  \\
 0.4--0.8   &   72      &  57.1    &     0.261   & 0.146  \\
 0.8--1.6   &   263     &  232     &     0.134   & 0.0265 \\
 1.6--3.2   &   956    &  925     &     0.0330  & 0.0457 \\
 3.2--6.4   &   3702    &  3650    &     0.0154  & 0.0132 \\
 6.4--12.8  &   14439   &  14200   &     0.0171  & 0.0281 \\\hline
\end{tabular}
\end{center}
\end{table}
   
Two important internal consistency checks on these results were performed:
using either synthetic \MgII\ absorbers with real LRGs or synthetic LRGs 
with real \MgII\ absorbers, we find no cross-correlation signal.

\subsection{Relative amplitude of cross- and auto-correlation}
\label{section:amplitude}

In order to constrain the amplitude of $w_{\rm ag}$ with respect to that of the
LRG--LRG auto-correlation $w_{\rm gg}$, we use the following estimator
with  the same galaxies used for the cross-correlation:
\begin{equation}
1+ w_{\rm gg}(r_\theta)= \frac{\rm GG}{\rm GR}\,. \label{eq:estimator:auto}
\end{equation}
GG is the total observed number of galaxy--galaxy pairs between
$r_\theta-\mathrm d r/2$ and $r_\theta+\mathrm d r/2$ and GR is the total
galaxy--random galaxy pairs, computed as before. 
 The filled triangles in Fig.~\ref{fig:xcorr:auto} show $w_{\rm gg}$. 
 The errors and the covariance matrix for $w_{\rm
gg}$ are computed using $N_{\rm jack}\!=\!10$ jackknife realisations.

The arguments in section~\ref{section:theory} require that both $w_{\rm ag}$ and $w_{\rm gg}$ have the
same slope $\gamma$. Therefore,  to constrain the  amplitude ratio $w_{\rm ag}/w_{\rm gg}$,
we first fitted a power law to $w_{\rm gg}(r_\theta)$, and used that as a
 model for ${w}_{\rm gg}$. We  will only use the scales
larger than $200h^{-1}{\rm \,kpc}$ in the rest of this paper
 in order to avoid possible cross-section effects
 (discussed at the end of this section).

First,   the model $\hat w_{\rm gg}(r_\theta)=A_{\rm gg} r_\theta^{\beta_{\rm gg}}$  for $w_{\rm gg}$
gives a best amplitude, $A_{\rm gg}$, at $1h^{-1}{\rm \,Mpc}$ and  slope $\beta_{\rm
gg}$ of \Aauto\ and \Bauto\ respectively~\footnote{The conversion of the amplitude, 
$A_{\rm gg}$, to the comoving length,
$r_0$, requires precise redshifts for the LRGs. As mentioned, only
$\sim\!160$ spectroscopic redshifts are available at present.
Nonetheless, a rough estimate   is $r_0\!=\!\rnot h^{-1}{\rm
\,Mpc}$ using the \MgII\ redshift distribution and
 assuming a Gaussian \dNdz\ with a FWHM $W_z=0.15$ for the  LRGs,
 and is  consistent with $6.3 \pm 0.5h^{-1}{\rm \,Mpc}$ found by
\citet{BrownM_03a}. }. The fitted power law
$\hat w_{\rm gg}(r_\theta)$ is shown by the dashed line in Fig.~\ref{fig:xcorr:auto}.

Then, from the following model for $w_{\rm ag}$,
 \begin{equation}
\hat w_{\rm ag}=a\times \hat w_{\rm gg},
\end{equation}
where $a$ is the amplitude ratio $A_{\rm ag}/A_{\rm gg}$,
we find that the best amplitude ratio 
\begin{equation}
a=\Arel \label{res:arel}
\end{equation}
 by minimizing
$\chi^2\propto [\mathbf{w}-\mathbf{\hat w}]^T \mathrm{COV}^{-1}[\mathbf{w}-\mathbf{\hat w}]\,,$
where   $\mathbf{w}$ and
$\mathbf{\hat w}$ are the vector data and model respectively and ${\rm
COV}^{-1}$ is the inverse of the covariance matrix,   calculated
using single value decomposition techniques \citep[see discussion in][]{BernsteinG_94a}.
This value is consistent with the fact that 5 of the 6 bins of $w_{\rm ag}$ at $r_\theta>200h^{-1}$~kpc
are below $w_{\rm gg}$ (Fig.~\ref{fig:xcorr:auto}).  Furthermore, the average  of
the bin ratios $w_{\rm ag}(r_\theta)/w_{\rm gg}(r_\theta)$ gives 0.85, close to $\Arel$.
 Fig.~\ref{fig:xcorr:auto} shows
both $\hat w_{\rm ag}$ (solid line) and $\hat w_{\rm gg}$ (dashed
line). 

For completeness, a power law fit to $w_{\rm ag}(r_\theta)$, 
i.e. $\hat w_{\rm ag}(r_\theta)=A_{\rm ag} r_\theta^{\beta_{\rm ag}}$,
gives $A_{\rm ag}=\Across$, the  fitted amplitude at $1h^{-1}{\rm
\,Mpc}$, , and $\beta_{\rm ag}\Bcross$, the slope.

Note that the relative amplitude, $a$, is free of
systematics from contaminants (e.g.~stars).
This is due to the fact that (1) we use the same galaxies for $w_{\rm ag}$ and
$w_{\rm gg}$, and (2) the estimators in equations~\ref{eq:estimator} and \ref{eq:estimator:auto}
are both $\propto 1/N_g$, where $N_g$ is the number of galaxies.
  Any contaminants will affect the cross- and
auto-correlation function in exactly the same way. We find that $a$ is also
robust under numerous different cuts and subsamples. For example, a more
restrictive star--galaxy separation [equation (\ref{LRG:selection2})] gives
consistent results. Similarly, $a$ is robust to a more stringent cut on the
redshift difference between the LRGs and \MgII\ absorbers than in equation
(\ref{LRG:selection3}). Finally, there is no significant variation in $a$
when excluding LRGs with larger or smaller SDSS magnitude errors.


If the \MgII\ cross-section radius ($R_\times$) is larger than
some of the radial bins, this will affect $w_{\rm ag}(r_\theta)$ and there
will be a redistribution of galaxies in the bins near $r_\theta\!\simeq\!R_\times$.
\citet{SteidelC_95b} constrained $R_\times$ to be $\sim70h^{-1}{\rm \,kpc}$ (comoving)
for absorbers with $W_{\rm r}^{\rm MgII}\!\geq\!0.3$\,\AA.
  In fact, we find that 
  (1) the first bin at
$r_\theta\!=\!50$--$100h^{-1}{\rm \,kpc}$ is higher than expected if
$w_{\rm ag}$ is a single power law extrapolated from the large scales (see
Table \ref{table:pairs});
(2) the second bin at
$r_\theta\!=\!100$--$200h^{-1}{\rm \,kpc}$ is negative and 2$\,\sigma$
below the fit in Fig.~\ref{fig:xcorr:auto}. The inset in
Fig.~\ref{fig:xcorr:auto} focuses on these two bins (indicated by the
vertical dotted lines) which are divided into two smaller sub-bins. 
The region from $r_\theta\!\sim\!100h^{-1}$ to $150h^{-1}{\rm \,kpc}$
(comoving) contains {\it no} galaxies. Note that since we used 
projected correlations $w(r_\theta)$,  this region corresponds to different angular
scales for the range of absorber redshifts, $0.4\!<\!z_{\rm
abs}\!<\!0.8$. We also find that this signature is present  even with less restrictive samples.
We speculate that given the results of \citet{SteidelC_95b},
this deficit of galaxies at $r_\theta\!=\!100$--$150h^{-1}{\rm \,kpc}$ is a signature of $R_\times$.
However, it  could be due to some other  physical process  that prevent pairs 
at that particular scale. 
 In a future paper, we will explore these hypotheses with larger data sets and simulations.

\section{Summary and Discussion}\label{section:summary}

 From the SDSS DR1,
we selected $\sim\!\ngalsim$ LRGs and 212 \MgII\ absorbers with $W_{\rm
r}^{\rm MgII}\!\geq\!1$\,\AA\ for \MgII\ \LL 2796 \& 2803 and $W_{\rm
r}^{\rm FeII}\!\geq\!0.5$\AA\ for \FeII\ \fa. 
We have examined the  clustering of these LRGs
around the \MgII\ absorbers with unprecedented statistics on small and large
scales ($0.05\!\la\!r_\theta\!\la\!13h^{-1}{\rm Mpc}$). 
The amplitude of the \MgII\--LRG cross-correlation
relative to that of the LRG--LRG auto-correlation is
 \Arelcorrsys, after applying a   correction of \overestimate\ per cent discussed   in
section~\ref{section:theory} and in the Appendix.
 The two error terms reflect  the statistical and systematic uncertainty,
respectively. By  adding the errors in quadrature,
\begin{equation}
a=\Arelcorr\,.\label{res:arelcorr}
\end{equation}
This corresponds to a  correlation length 
$r_{\rm 0,ag}=(a)^{1/1.8}r_{\rm 0,gg}=5.04^{+0.62}_{-0.63} h^{-1}$~Mpc for $w_{\rm ag}$, and  
$r_{\rm 0,aa}=(a)^{2/1.8}r_{\rm 0,gg}=4.04^{+0.78}_{-0.78} h^{-1}$~Mpc for the 
\MgII--\MgII\ autocorrelation $w_{\rm aa}$, where we
 used the minimum auto-correlation length $r_{\rm 0,gg}=6.3 \pm 0.5 h^{-1}{\rm \,Mpc}$ of LRGs from
\citet{BrownM_03a}.

Within the context of hierarchical galaxy formation 
[equations (\ref{eq:biasCDM}--\ref{eq:biascross})],
equation~\ref{res:arelcorr} implies that our \MgII\ absorbers have  halo masses  4--20 times smaller
than the  LRG mininum-mass ($\LRGminmass$\,\msun), or \LRGmassrangemin.
For $10^{13}$\,\msun\ LRG halos, the \MgII\ absorbers have halos of \LRGmassrange.

Is our mass constraint of  \MgII\ absorbers in agreement with the results of \citet{SteidelC_94a} 
who found that \MgII\ absorbers with $W_{\rm
r}^{\rm MgII}\!\geq\!0.3$\,\AA\ are associated with $\sim\!0.7L_B^*$  
galaxies? Our mass measurement appears broadly consistent with those results
given that  $\sim L_B^*$ galaxies have halos of mass $\sim\!10^{12}$\msun. Furthermore,
the expected amplitude ratio is $\sim\!0.70$, close to our $a\!=\!0.67$.
The  expected amplitude ratio is found
assuming that the correlation length does not evolve from $z=0.5$
to $z=0$ and using the local correlation of early and late type galaxies.
At $z=0$, \citet{ShepherdC_01a} found that the early- and
late-type galaxy auto-correlation lengths were
$r_0\!=5.45{\rm \,Mpc}$   and $3.95$ respectively.  
 \citet{BudavariT_03a} found $r_0\!=6.5{\rm \,Mpc}$   and $4.5$ respectively.
 Assuming  $\gamma\!=\!1.8$ for both of these auto-correlations, then from
equations \ref{eq:biasCDM} \& \ref{eq:biascross} one expects the
late--early cross-correlation amplitude to be
$(3.95/5.45)^{1.8/2}\!\simeq\!0.74$ \citep[0.72 for][]{BudavariT_03a}
times that of the auto-correlation.

Note that there are important differences between our \MgII\ sample and that of
\citet{SteidelC_94a}. Firstly, our
larger equivalent width threshold, $W_{\rm r}^{\rm MgII}\!\geq\!1$\,\AA,
will preferentially select systems with a larger velocity dispersion over
the absorption components. Thus, our sample is potentially
biased towards more massive halos. Secondly, our \MgII\ sample will be dominated by DLAs:
\citet{RaoS_00a} find that $\sim\!50$\,per cent of systems with $W_{\rm
r}^{\rm MgII}\!\geq\!0.5$\,\AA\ and $W_{\rm r}^{\rm FeII}\!\geq\!0.5$\,\AA\
are DLAs. \citet{NestorD_03a} use $W_{\rm r}^{\rm MgII}\!\geq\!1$\,\AA\
to select a larger proportion of DLAs.
 
 It should be emphasized that this method (i.e. measuring a correlation ratio)
 has the following advantages:
 (i) it is free of systematics from contaminants (e.g.~stars),
 (ii) it  does not require  knowledge of the true width of the redshift distribution,
and (iii) it constrains the masses of the \MgII/DLA
host-galaxies in a statistical manner without directly identifying them.
Thus, with a sample of confirmed DLAs which are not selected on the basis
of \MgII\ line-strength, one should be able to derive the mean mass of the
DLA host-galaxies with only relatively shallow wide-field imaging. This
could help establish the relative proportions of low- and high-luminosity
contributions to DLA host-galaxies. This topic is currently under some
debate (e.g.~compare \citealt{RaoS_03a} and \citealt{ChenH-W_03a}).

\section*{Acknowledgments}

We thank Brice M\'enard for many helpful discussions;
 Paul Hewett and Max Pettini for
their comments; D. Croton  for providing the catalog of the GIF2 simulation.
 We also thank the anonymous referee for a swift review that led to an improved analysis.
  This work was supported by the European Community Research and Training Network `The
Physics of the Intergalactic Medium'.  
MTM is grateful to PPARC for support
at the IoA. Funding for the Sloan
Digital Sky Survey (SDSS) has been provided by the Alfred P. Sloan
Foundation, the Participating Institutions, the National Aeronautics and
Space Administration, the National Science Foundation, the U.S. Department
of Energy, the Japanese Monbukagakusho, and the Max Planck Society.


\appendix
\section[On correlation functions]{On correlation functions\protect\footnote{This appendix presents 
and organizes previously published results, and
is therefore not part of the published text.}} 

\label{section:definitions}

It may seem that taking the ratio between the cross- and auto-correlation
is inappropriate since the former is based on absorbers with spectroscopic
(i.e.~accurate) redshifts and a sample of galaxies with photometric
redshifts  (accurate only to $\sigma_z\simeq 0.1$), while the
latter comprises only galaxies with photometric redshifts. In this paper,
we have measured the projected correlation function $w_{\rm p}(r_\theta)$.
For a given field (with one absorber) with galaxies distributed with \dNdz,
one may think that  the auto-correlation
 is proportional to $\int \left(\dNdz\right)^2 \mathrm d z$
while the cross-correlation is proportional to $\int \left(\dNdz\right)^{1}
\mathrm d z$. Thus, at first glance,  their ratio is therefore
not very useful. Below we show
the situation to be not so trivial.
 
First, some definitions and results that will be useful later. For a 3D
correlation function $\xi(r)=(r/r_{0})^{-\gamma}$, the projected
correlation function $w_{\rm p}(r_{\rm p})$ is \citep{DavisM_83a}:
\begin{eqnarray}
 w_{\rm p}(r_{\rm p})&=&\int_\infty^\infty \mathrm d y \; \xi(r_{\rm p},y)
 		=\int_\infty^\infty \mathrm d y \; \xi(\sqrt{r_{\rm
 		p}^2+y^2}) \nn \\ &=& (r_{\rm p})^{1-\gamma}\; r_{0}^\gamma
 		\; H_\gamma \label{croft}
\end{eqnarray}
where $\xi(r_{\rm p},y)$ is the 3D correlation function decomposed along
the line of sight $y$ and on the plane of the sky $r_{\rm p}$, i.e. $r^2 =
y^2 + r_{\rm p}^2$. $H_\gamma$ is in fact the Beta function
$B(a,b)=\int_0^1 t^{a-1}\;(1-t)^{b-1}\;\mathrm d t$ evaluated with $a=1/2$
and $b=(\gamma-1)/2$,
i.e.~$H_\gamma=B(\frac{1}{2},\frac{\gamma-1}{2})=\Hgamma$.

In appendix C of \citet{AdelbergerK_03a}, one finds the expected number of
neighbours between $r_\theta-\mathrm d r/2$ and $r_\theta + \mathrm d r /2$
within a redshift distance $|\Delta_z|\!<\!r_{\rm z}$:
\begin{eqnarray}
w_{\rm p}(r_{\theta}\!<\!r_{\rm z})&=&\frac{1}{r_{\rm z}}\int_0^{r_z}
 		\mathrm d l \; \xi(\sqrt{r_{\theta}^2+l^2}) \nn \\ &=&
 		\frac{1}{2 r_z} (r_{\theta})^{1-\gamma}\; r_{0}^\gamma \;
 		H_\gamma\; I_x(\frac{1}{2},\frac{\gamma-1}{2})
		\label{adelberger}
\end{eqnarray}
where $x=r_{\rm z}^2/(r_{\rm z}^2+r_\theta^2)$ and $I_x$ is the
incomplete Beta function $B_x(a,b)=\int_0^x t^{a-1}\;(1-t)^{b-1}\;\mathrm d
t$ normalized by ${B(a,b)}$: $I_x(a,b)\equiv B_x(a,b)/B(a,b)$.

Many papers \citep{PhillippsS_78a,PeeblesP_93a,BudavariT_03a} have shown
that the angular correlation function is
\begin{eqnarray}
w(\theta)=(\theta)^{1-\gamma}\; r_{0}^\gamma \; H_\gamma \times
\int_0^\infty \mathrm d z \left(\dNdz\right)^2 g(z)^{-1} f(z)^{1-\gamma}
\label{budavari}
\end{eqnarray}
where $g(z)=\mathrm d r/\mathrm d z=c/H(z)$ and $f(z)=D_c(z)$ is the
comoving line-of-sight distance to redshift $z$, i.e.~$D_c(z)=\int_0^z
\mathrm d t \frac{c}{H(t)}$.

Equation~\ref{budavari} can be derived from the definitions of the angular
and 3D correlation functions, $w(\theta)$ and $\xi(r)$
\citep[e.g.][]{PhillippsS_78a}. We reproduce the derivation here and extend
it to projected auto- and cross-correlation functions. The probabilities of
finding a galaxy in a volume d$V_1$ and another in a volume d$V_2$ at a
distance $ r=|\mathbf r_2\!-\!\mathbf r_1|$, along two lines of sight separated by $\theta$ are
\begin{eqnarray}
\mathrm d P(\theta)&=&{\cal N}^2\; \mathrm d \Omega_1 \mathrm d \Omega_2
[1+w(\theta)] \label{definition:angular}\,\hbox{or}\\
\mathrm d P(r) &=& n(z)^2 \;\mathrm d V_1 \mathrm d V_2 [1+\xi(r)]
\label{definition:xi}
\end{eqnarray}
where $\cal N$ is the number of galaxies per solid angle, i.e.~$\mathrm d
N/\mathrm d \Omega$, and $n(z)$ is the number density of galaxies, which
can be a function of redshift. Given that ${\cal N}=\frac{1}{\mathrm d
\Omega} \int \, n(z) \mathrm d V(z)$ and that $\mathrm d V= f^2(z) g(z)
\mathrm \; d \Omega \mathrm d z$, ${\cal N}\equiv \int \mathrm d z\;\dNdz =
\int \mathrm d z \; n(z) f^2(z)g(z)$.

To relate $w(\theta)$ and $\xi(r)$, one needs to integrate equation
\ref{definition:xi} over all possible lines-of-sight separated by $\theta$
(i.e. along $z_1$ and $z_2$) and equate it with
equation \ref{definition:angular}:
\begin{eqnarray}
{\cal N}^2[1+w(\theta)]&=&\int_0^\infty\mathrm d z_1 f(z_1)^2g(z_1)n(z_1)\cdot
\nn \\ && \int_0^\infty d z_2 f(z_2)^2g(z_2)n(z_2)[1+\xi(r_{12})] \;
. \label{eq1}
\end{eqnarray}
In the regime of small angles, the distance $r_{12}$ (in comoving Mpc) can
be approximated by:
\begin{eqnarray}
r_{12}^2&=&r_1^2+r_2^2-2r_1r_2 \cos \theta \nn\\ &\simeq&(r_1-r_2)^2+r^2
	\theta^2 \quad \hbox{with}\; r=\frac{r_1+r_2}{2} \nn \\
	&\simeq&(g(z)(z_1-z_2))^2+f(z)^2 \theta^2 \quad \hbox{with}\;
	z=\frac{z_1+z_2}{2} \nn\\ &\simeq&g(z)^2y^2+f(z)^2 \theta^2 \quad
	\hbox{with}\; y={z_1-z_2}\, . \label{smallangle}
\end{eqnarray}
Changing variables in equation~\ref{eq1} from ($z_1,z_2$) to ($z,y$),
assuming the the major contribution is from $z_1\simeq z_2$ and using
equation~\ref{smallangle}, the angular correlation function is
\begin{eqnarray}
w(\theta)=\frac{ \int_0^\infty d z f(z)^4g(z)^2n(z)^2\int_{-\infty}^\infty
\mathrm d y \xi(\sqrt{f(z)^2\theta^2+g(z)^2y^2})}{\left[ \int_0^\infty d z
f^2(z)g(z)n(z)\right ]^2}\,. \label{eq:angular}
\end{eqnarray}
Changing variables to $l=g(z)y$, using equation~\ref{croft} and using a
normalized redshift distribution, i.e. $\int \mathrm d z \;\dNdz=1$,
equation~\ref{eq:angular} becomes
\begin{eqnarray}
w(\theta)= \int_0^\infty d z \left(\dNdz \right)^2 g(z)^{-1} \times (f(z)
\theta)^{1-\gamma}\; r_{0}^\gamma \; H_\gamma \label{intermadiate}
\end{eqnarray}
which leads to equation~\ref{budavari} \citep[equation~9
in][]{BudavariT_03a} and is one version of Limber's equations.

In this paper, we measured the projected auto-correlation of the LRGs,
$w_{\rm gg}(r_\theta)$, where $r_\theta=f(z)\theta$~\footnote{In general
this should be $D_A (1+z) \theta$ where $D_A$ is the angular distance.  For
a flat universe, $D_A (1+z)=D_M=Dc=f(z)$ where $D_M$ is the comoving
transverse distance, using D. Hogg's notations \citep{HoggD_99a}.}.
Following the same steps as above with $r_\theta$ instead of $\theta$, and
$\mathrm d V=(\mathrm d r_\theta)^2 g(z) \mathrm d z$, $w_{\rm
gg}(r_\theta)$ is:
\begin{eqnarray}
w_{\rm gg}(r_\theta)= r_\theta^{1-\gamma}\; r_{\rm 0,gg}^\gamma \; H_\gamma
\int_0^\infty d z \left(\dNdz \right)^2 g(z)^{-1}
\label{auto:result}
\end{eqnarray}
 
In the case of the projected cross-correlation, $w_{\rm ag}(r_\theta)$, the
conditional probability of finding a galaxy in the volume $\mathrm d V_2$
given that there is an absorber at a known position $\mathbf r_1$ is, by
definition \citep[e.g.][]{EisensteinD_03a},
\begin{eqnarray}
\mathrm d P(2|1)(\theta)&=&{\cal N}_{\rm g}\;  \mathrm d \Omega_2 [1+w_{\rm p}(r_\theta)]  \label{definition2:angular} \\
\mathrm d P(2|1)(r) &=& n_{\rm g}(z) \; \mathrm d V_2 [1+\xi(r)] \label{definition2:xi}\,.
\end{eqnarray}
Using the same approximations (equation~\ref{smallangle}) and one integral
along the line of sight $z_2$  (keeping the absorber at $z_1$), one finds
that the projected cross-correlation is:
\begin{eqnarray}
w_{\rm ag}(r_\theta)&=& \int_0^\infty d z_2 f(z_2)^2g(z_2)n(z_2) \xi(r_{12}) \nn \\
&=& \int_0^\infty d z \left(\dNdz \right)  \xi(\sqrt{r_\theta^2+g(z)^2(z_1-z_2)^2})\nn\\
&=& \int_0^\infty d y \,g(z) \left(\dNdz \right) g(z)^{-1} \xi(\sqrt{r_\theta^2+g(z)^2y^2})\nn\\
&=& \int_0^\infty d l  \frac{\mathrm d N}{\mathrm d l} \xi(\sqrt{r_\theta^2+l^2})\label{cross:result}\\
&\simeq&\frac{1}{W_z}\times (r_\theta)^{1-\gamma}\; r_{\rm 0,ag}^\gamma \; H_\gamma \,,
\end{eqnarray}
where we approximated \dNdz\ with a normalized top-hat of width $W_z=2r_z$,
used equation~\ref{adelberger}, and the fact that $I_x\simeq 1$ since
$x\simeq 1$ for a typical width $W_z$ of 200$h^{-1}$~Mpc (Section~2.2).
Thus, as one would have expected, the cross-correlation is inversely
proportional to the width of the galaxy distribution.  Naturally, in
equation~\ref{definition2:angular} and \ref{definition2:xi}, the redshift
of galaxy 1 (i.e.~the absorber) is assumed to be known with good
precision. If the absorber population had poorly known redshifts, one would
need to add an integral to equation~\ref{cross:result}, washing out the
cross-correlation signal further. This is not an issue for our \MgII\
absorbers.

For the projected auto-correlation (equation~\ref{auto:result}), if one
approximates \dNdz\ by a top-hat function of width $W_z$, then
\begin{eqnarray}
w_{\rm gg}(r_\theta)
&=& (r_\theta)^{1-\gamma}\; r_{\rm 0,ag}^\gamma \; H_\gamma  \times \int_0^\infty \mathrm d z \, g(z) \left(\dNdz\right)^2 g(z)^{-2}
  \nn \\
&=& (r_\theta)^{1-\gamma}\; r_{\rm 0,gg}^\gamma \; H_\gamma  \times \int_0^\infty \mathrm d l \left(\frac{\mathrm d N}{\mathrm d l}\right)^2   
  \nn \\
&\simeq& \left( \frac{1}{W_z}\right )^2 \; W_z \times (r_\theta)^{1-\gamma}\; r_{\rm 0,gg}^\gamma \; H_\gamma \,,
\end{eqnarray}
which shows that the auto-correlation depends on the redshift distribution
of the galaxies in the same way as the cross-correlation, i.e. $\propto
1/W_z$. The reason for this is
 that the redshift distribution \dNdz\ has a very different role with respect to the correlation functions,
 which can be seen by comparing equations~\ref{auto:result} and \ref{cross:result}.
It is this   very different role that leads to the same  $1/W_z$ dependence.

  The above considerations were for one absorber and can be easily extended for
many absorbers, since the projected correlations are measured at the same
scales (by definition): $\overline w_{\rm p}(r_\theta)=\frac{1}{N_{\rm
a}}\sum_i^{N_{\rm a}} w_{{\rm p},i}(r_\theta)$, where $w_{{\rm p},i}$ is
the projected correlation function for one field and $N_{\rm a}$ is the
number of absorbers (or fields).

In the case of a Gaussian redshift distribution \dNdz, the ratio of cross-
and auto-correlations may not be exactly unity.  Using Mock galaxy samples
\citep[from the GIF2 collaboration,][]{GaoL_04a} selected in a redshift
slice of width, $W_z$, equal to their artificial Gaussian redshift errors
$\sigma_z$, we find that the cross-correlation is overestimated by
\overestimate\,per cent.  Quite importantly, this correction factor is
independent of the width of the redshift distribution as long as
$\sigma_z\simeq W_z$  or as long as it is Gaussian.  This implies that the
ratio of the correlation functions ($w_{\rm ag}/w_{\rm gg}$) will be
insensitive to errors in photometric redshifts.


\bsp_small

\label{lastpage}

\end{document}